\PassOptionsToPackage{unicode}{hyperref}
\documentclass[a4paper,UKenglish,cleveref, autoref, thm-restate]{lipics-v2021}
\nolinenumbers

\hypersetup{
    colorlinks=true,
    filecolor=magenta,      
    urlcolor=blue,
}

\usepackage{subcaption}

\usepackage{array}
\newcolumntype{L}{>{\centering\arraybackslash}m{3cm}}

\long\def\ignore#1{}

\title{Compressing and Indexing Aligned Readsets}

\author{Travis Gagie}{Dalhousie University, Canada}{travis.gagie@gmail.com}{https://orcid.org/0000-0003-3689-327X}{Funded by NSERC Discovery Grant RGPIN-07185-2020, NIH R01HG011392 and NSF IIBR 2029552.}

\author{Garance Gourdel}{IRISA - Inria Rennes - Université Rennes 1 - ENS, France}{garance.gourdel@irisa.fr}{}{}

\author{Giovanni Manzini}{University of Pisa, Italy}{giovanni.manzini@unipi.it}{https://orcid.org/0000-0002-5047-0196}{Supported by the Italian MIUR PRIN project 2017WR7SHH.}

\authorrunning{T. Gagie, G. Gourdel and G. Manzini}

\Copyright{Travis Gagie, Garance Gourdel and Giovanni Manzini}



\supplement{\url{https://github.com/fnareoh/Big_XBWT}}

\acknowledgements{Many thanks to Jarno Alanko and Uwe Baier for their XBWT-construction software, and to Diego D\'iaz, Richard Durbin, Filippo Geraci, Giuseppe Italiano, Ben Langmead, Gonzalo Navarro, Pierre Peterlongo, Nicola Prezza, Giovanna Rosone, Jared Simpson, Jouni Sir\'en and Jan Studen\'y for helpful discussions.}

\EventEditors{Alessandra Carbone and Mohammed El-Kebir}
\EventNoEds{2}
\EventLongTitle{21st International Workshop on Algorithms in Bioinformatics (WABI 2021)}
\EventShortTitle{WABI 2021}
\EventAcronym{WABI}
\EventYear{2021}
\EventDate{August 2--4, 2021}
\EventLocation{Virtual Conference}
\EventLogo{}
\SeriesVolume{201}
\ArticleNo{13}

\begin{document}

\maketitle

\begin{abstract}

Compressed full-text indexes are one of the main success stories of bioinformatics data structures but even they struggle to handle some DNA readsets.  This may seem surprising since, at least when dealing with short reads from the same individual, the readset will be highly repetitive and, thus, highly compressible.  If we are not careful, however, this advantage can be more than offset by two disadvantages: first, since most base pairs are included in at least tens reads each, the uncompressed readset is likely to be at least an order of magnitude larger than the individual's uncompressed genome; second, these indexes usually pay some space overhead for each string they store, and the total overhead can be substantial when dealing with millions of reads.

The most successful compressed full-text indexes for readsets so far are based on the Extended Burrows-Wheeler Transform (EBWT) and use a sorting heuristic to try to reduce the space overhead per read, but they still treat the reads as separate strings and thus may not take full advantage of the readset's structure.  For example, if we have already assembled an individual's genome from the readset, then we can usually use it to compress the readset well: e.g., we store the gap-coded list of reads' starting positions; we store the list of their lengths, which is often highly compressible; and we store information about the sequencing errors, which are rare with short reads.  There is nowhere, however, where we can plug an assembled genome into the EBWT.

In this paper we show how to use one or more assembled or partially assembled genome as the basis for a compressed full-text index of its readset.  Specifically, we build a labelled tree by taking the assembled genome as a trunk and grafting onto it the reads that align to it, at the starting positions of their alignments.  Next, we compute the eXtended Burrows-Wheeler Transform (XBWT) of the resulting labelled tree and build a compressed full-text index on that. Although this index can occasionally return false positives, it is usually much more compact than the alternatives.
Following the established practice for datasets with many repetitions, we compare different full-text indices by looking at the number of runs in the transformed strings. For a human Chr19 readset our preliminary experiments show that eliminating separators characters from the EBWT reduces the number of runs by 19\%, from 220 million to 178 million, and using the XBWT reduces it by a further 15\%, to 150 million.

\end{abstract}

\section{Introduction}
\label{sec:introduction}

The FM-index~\cite{ferragina2005indexing} is an important data structure in both combinatorial pattern matching and bioinformatics.  Its most important application so far has been in standard short-read aligners --- Bowtie~\cite{langmead2009ultrafast,langmead2012fast} and BWA~\cite{li2009fast} have together over 70 thousand citations and are used every day in clinics and research labs worldwide --- but it has myriad other uses and more are still being discovered.  Just within computational genomics, FM-indexes have been generalized from single strings to collections of strings for tools such as BEETL~\cite{Beetl}, RopeBWT~\cite{ropebwt2} and Spring~\cite{spring}, to de Bruijn graphs for tools such as BOSS~\cite{BOSS}, VARI~\cite{muggli2017succinct} and Rainbowfish~\cite{almodaresi2017rainbowfish}, and to graphs for tools such as vg~\cite{garrison2018variation}. Recent breakthroughs~\cite{gagie2020fully} mean we can now scale FM-indexes to massive but highly repetitive pan-genomic datasets for a new generation of tools~\cite{kuhnle2020efficient}.

As genomic datasets grow exponentially (from the Human Genome Project to the 1000 Genomes Project and the 100K Genomes Project) and standards for sequencing coverage increase (from less than 10x a few years ago to 30x and 50x now and over 100x for some applications), an obvious question is whether and how the recent breakthroughs in FM-indexing of repetitive datasets can be turned into comparable advances in indexing readsets, so more researchers can efficiently mine them for biomedical insights. For example, extrapolating from previous experiments~\cite{kuhnle2020efficient}, it should be possible to index both haplotypes from  2705 individuals in less than 100 GB of RAM.  In contrast, the readset from the final phase of the 1000 Genomes Project consisted of reads from 2705 individuals and was released as a 464 GB Burrows-Wheeler Transform (BWT)~\cite{dolle2017using}, which is beyond the resources of most labs to process.
This almost five-fold increase (from 100 to 464 GB) seems reasonable, given the range of lengths and the error rate of short-read sequencing technologies, but those reads were trimmed and error-corrected before their BWT was computed, making that increase harder to justify and thus a target for improvement.  Although experimenting with that particular readset is beyond the scope of this paper, since it occupies 87 TB uncompressed, we expect the insights and techniques we develop here will eventually be useful in software able to handle efficiently inputs of that scale.

{Recent results on FM-indexing repetitive datasets~\cite{gagie2020fully} have shown that the index performance depends on the number of runs in the transformed sequence, where a run is a maximal non-empty unary substring. For example, if the BWT of a dataset of (uncompressed) size $n$ has $r$ runs, we can design an FM-index of size $O(r \log\log n)$ supporting the count and locate operations in optimal linear time. Hence, if a BWT variant produces a transformed string with a smaller number of runs, the resulting index will be smaller and equally fast.} The na\"ive approach to FM-indexing readsets is to concatenate the reads with copies of a separator character between them, and FM-index the resulting single string.  However, computing the BWT of such a long string is a challenge and each separator character causes several runs in that BWT.  The most competitive indexes for readsets are based on Mantaci et al.'s~\cite{MANTACI2007298} Extended Burrows Wheeler Transform, which is also easier to build for readsets.  
The first index for readsets based on the EBWT was BEETL~\cite{Beetl}, followed by RopeBWT~\cite{ropebwt2}; recently the EBWT has been used also by the Spring compressor~\cite{spring} specialized for FASTQ reads. BEETL and RopeBWT use explicit separator characters but such characters could be replaced by bitvectors marking positions at the ends of reads. %the experiments in this paper indicate that the use of bitvectors is indeed worthwhile. 

BEETL and RopeBWT use a heuristic to reduce the number of runs in the EBWT: they conceptually put the separator characters at the ends of reads into the co-lexicographic order (lexicographic order on the reverse string, also referred to as reverse lexicographic order) of the reads, so that the final characters or reads with similar suffixes are grouped together in the EBWT.  This often works surprisingly well but in the worst case it cannot make up for the lack of context for sorting those characters into their places in the EBWT.  Our proposal in this paper is to graft the reads onto their assembled genome, or a reference genome to which they align well, and index the resulting labelled tree with Ferragina et al.'s~\cite{ferragina2009compressing} XBWT.
To this end we assume that we know how the reads align to the assembled/reference genome: this is not an unreasonable assumption since alignment is the initial step of any readset analysis.

In order to implement our idea we have to overcome a significant hurdle: as the coverage increases so does the amount of raw data produced by a single NGS experiment. Although the high coverage implies that the data is highly compressible, the actual compression process, ie the construction and the compression of the XBWT, must be done partially in externally memory since the input will be usually much larger than the available RAM. Another contribution of the paper is therefore the adaptation of the prefix-free parsing (PFP) technique~\cite{boucher2019prefix} to the construction of the XBWT. PFP has been proposed for the construction of BWTs of collections of similar genomes: the initial parsing phase is able to compress the input maintaining enough information to compute the BWT working on the compressed representation. In this paper we adapt PFP to readsets, taking care also of the ``grafting'' of the single reads to the reference/assembled genome.
Given a pattern $P$, our index could answer $count(P)$ and $locate(P)$ queries which report respectively the number of positions where $P$ occurs and the list of positions where $P$ occurs. The main drawback to our index, apart from taking one or more assembled or partially assembled genomes as a base, is that it can return a false-positive in the $count$ operation when an occurrence of a pattern starts in the trunk of an alignment tree and ends in a branch.  In other words, the index can report a match that is not completely contained within a read but would be if we padded the read on the left with enough characters copied from just before where it aligns. In a locate operation false-positives could be identified, but this operation is much slower. Even this is not entirely bad, however, and it is conceivable this bug could sometimes be a feature.
The analysis of those false positive and the size of the bit vectors marking the end of reads is left as future work.

The rest of the paper is organized as follows. In Section~\ref{sec:concepts} we first describe the BWT and FM-indexes, then the EBWT and XBWT and the concept of Wheeler graph that unifies them. 
In Section~\ref{sec:main} we introduce our idea for indexing aligned readsets with the XBWT {and we prove some theoretical results supporting it}.  In Section~\ref{sec:pfp} we describe how we adapt PFP to indexing readsets, which allows us to experiment with larger files than would otherwise be possible with reasonable resources. In Section~\ref{sec:practice} we present our experimental results showing that applying the XBWT to index readsets works well in practice as well as in theory. Finally, we outline in Section~\ref{sec:JST} how our study of storing reads with the XBWT may improve the space usage of the hybrid index~\cite{ferrada2014hybrid,ferrada2018hybrid,gagie2015searching}.

\section{Concepts} \label{sec:concepts}

{For a better understanding of the problem context, we give a succinct description of the second generation sequencing technique.} Most publicly available readsets are from Illumina sequencers~\cite{kodama2012sequence} which rely on sequencing by synthesis.  For this process, millions or billions of single-stranded snippets of DNA called templates are deposited onto a slide and amplified into clusters of clones.  In each sequencing cycle we learn one base of each template: we add DNA polymerase and specially terminated bases; the polymerase attaches a terminated base to each strand, complementary to the next base in the strand; we shine a light on the slide and the terminated bases glow various colours; we take a photo and note the colour of each cluster; and finally, we treat the slide to remove the terminators.  Sometimes, however, one of the added bases is not correctly terminated, so the polymerase attaches first it and then another base to a strand in some cluster; that strand is then out of step with the rest of the cluster, and the cluster will have a mix of colours in the photos for subsequent sequencing cycles.  As we go through more and more sequencing cycles, more strands tend to fall out of step, resulting in less reliable results.  (For futher discussion we refer the reader to, e.g., Langmead's lecture on this topic~\cite{LangmeadVideo}.)  This tendency means sequencing by synthesis has an asymmetric error profile, with errors more likely towards the ends of the reads.  It follows that sequencing errors tend to be near the end of the reads: our index is designed to take advantage of this feature (see Theorem~\ref{thm:XBWTerr}).

\subsection{BWT and FM-index}
\label{subsec:BWT}

The Burrows-Wheeler Transform (BWT)~\cite{burrows1994block} of a string $S$ is a permutation of the characters in $S$ into the lexicographic order of the suffixes that immediately follow them, considering $S$ to be cyclic.  For example, as shown on the left in Figure~\ref{fig:dollars}, the BWT of {\tt GATTAGATACAT\$} is {\tt TTTCGGAA\$AATA}, assuming {\tt \$} is a special end-of-string symbol lexicographically smaller than all other characters.  Because the BWT groups together characters that precede similar suffixes, it tends to convert global repetitiveness into local homogeneity: e.g., for any string $\alpha$, the BWT of $\alpha^t$ consists of $|\alpha|$ unary substrings of length $t$ each; even the BWT in our example has length 13 but consists of only 8 maximal unary substrings (called runs).  This property led Burrows and Wheeler to propose the BWT as a pre-processing step for data compression and Seward~\cite{seward1996bzip2} used it as the basis for the popular {\tt bzip2} compression program.

The BWT is also the basis for the FM-index~\cite{ferragina2005indexing}, one of the first and most popular compressed indexes, which is essentially a rank data structure over the BWT combined with a suffix-array sample.  The FM-index is an important data structure in combinatorial pattern matching and bioinformatics, and is itself the basis for popular tools such as Bowtie~\cite{langmead2009ultrafast,langmead2012fast} and BWA~\cite{li2009fast} that align DNA reads to reference genomes.  We refer the reader to Navarro's~\cite{navarro2016compact} and M\"akinen et al.'s~\cite{makinen2015genome} textbooks for detailed discussions of how FM-indexes are implemented and used for read alignment.

\begin{figure}[t]
\begin{center}
\begin{tabular}{c@{\hspace{20ex}}c}
\begin{tabular}{rc}
   & $F$ \hspace{11ex} $L$\\
\hline
 0 & \tt \$GATTAGATACAT\\
 1 & \tt ACAT\$GATTAGAT\\
 2 & \tt AGATACAT\$GATT\\
 3 & \tt AT\$GATTAGATAC\\
 4 & \tt ATACAT\$GATTAG\\
 5 & \tt ATTAGATACAT\$G\\
 6 & \tt CAT\$GATTAGATA\\
 7 & \tt GATACAT\$GATTA\\
 8 & \tt GATTAGATACAT\$\\
 9 & \tt T\$GATTAGATACA\\
10 & \tt TACAT\$GATTAGA\\
11 & \tt TAGATACAT\$GAT\\
12 & \tt TTAGATACAT\$GA
\end{tabular} &
\begin{tabular}{rl@{\hspace{6ex}}rl}
   & $F$ \hspace{3.8ex} $L$ && $F$ \hspace{3.8ex} $L$\\
\hline
 0 & \tt \$ATACAT   & 16 & \tt ATTA\$\ G\\
 1 & \tt \$GATA\ C  & 17 & \tt C\$GAT\ A\\
 2 & \tt \$GATT\ A  & 18 & \tt CAT\$ATA\\
 3 & \tt \$TAGATA   & 19 & \tt GA\$TT\ A\\
 4 & \tt \$TTAG\ A  & 20 & \tt GATA\$TA\\
 5 & \tt A\$GAT\ T  & 21 & \tt GATAC\ \$\\
 6 & \tt A\$TAGAT   & 22 & \tt GATTA\ \$\\
 7 & \tt A\$TTA\ G  & 23 & \tt T\$ATACA\\
 8 & \tt AC\$GA\ T  & 24 & \tt TA\$GA\ T\\
 9 & \tt ACAT\$AT   & 25 & \tt TA\$TAGA\\
10 & \tt AGA\$T\ T  & 26 & \tt TAC\$G\ A\\
11 & \tt AGATA\$T   & 27 & \tt TACAT\$A\\
12 & \tt AT\$ATAC   & 28 & \tt TAGA\$\ T\\
13 & \tt ATA\$TAG   & 29 & \tt TAGATA\$\\
14 & \tt ATAC\$\ G  & 30 & \tt TTA\$G\ A\\
15 & \tt ATACAT\$   & 31 & \tt TTAGA\ \$
\end{tabular}
\end{tabular}
\caption{The matrices whose rows are the lexicographically sorted rotations of {\tt GATTAGATACAT\$} {\bf (left)} and of {\tt GATTA\$}, {\tt TTAGA\$}, {\tt TAGATA\$}, {\tt GATAC\$} and {\tt ATACAT\$} {\bf (right)}.  The BWT and EBWT are {\tt TTTCGGAA\$AATA} and {\tt TCAAATTGTTTTCGG\$GAAAA\$\$ATAAAT\$A\$} with 8 and 19 runs, respectively.}
\label{fig:dollars}
\end{center}
\end{figure}

\subsection{EBWT}
\label{subsec:EBWT}

Although alignment against one or more reference genomes remains a key task in bioinformatics, there is growing interest in compressed indexing of sets of reads~\cite{dolle2017using,kayegenome}.  The FM-index plays a central role here too: Mantaci et al.~\cite{mantaci2007extension} generalized the BWT to the Extended BWT (EBWT), which applies to collections of strings, and then Cox et al.~\cite{bauer2013lightweight,cox2012large,janin2014beetl} used an FM-index built on the EBWT in their index BEETL for readsets.  The same construction was also used in subsequent indexes for readsets, such as RopeBWT~\cite{ropebwt2} and Spring~\cite{spring}.

The EBWT of a collection of strings is a permutation of the characters in those strings into the lexicographic order of the suffixes that immediately follow them, considering each string to be cyclic.  For example, as shown on the right in Figure~\ref{fig:dollars}, the EBWT of {\tt GATTA\$}, {\tt TTAGA\$}, {\tt TAGATA\$}, {\tt GATAC\$} and {\tt ATACAT\$} is {\tt TCAAATTGTTTTCGG\$GAAAA\$\$ATAAAT\$A\$}.  When we see the BWT and EBWT as permutations of characters, the BWT of a single string has a single cycle, whereas the EBWT of a collection of strings has a cycle for each string.  This means it is easier to build the EBWT and update it when a string is added or deleted, than to build and update the BWT of the concatenation of the collection with the strings separated by copies of a special character.  We refer the reader to Egidi et al.'s~\cite{egidi2019external,tcs/EgidiM20} and D\'iaz-Dom\'inguez and Navarro's~\cite{DNdcc21.3} recent papers for descriptions of efficient construction and updating algorithms.

Despite its benefits, the EBWT sometimes does not take full advantage of its input's compressibility.  In our example, as Figure~\ref{fig:dollars} shows, even though all the strings in the collection are substrings of {\tt GATTAGATACAT\$} with copies of {\tt \$} appended to them, their EBWT has more than twice as many runs as its BWT.  As a heuristic for reducing the number of runs, and thus reducing BEETL's space usage, Cox et al.\ suggested considering the lexicographic order of the copies of {\tt \$} to be the strings' co-lexicographic order.  This does not help in cases such as our example, however, for which the EBWT still has 19 runs even with that ordering.  Bentley et al.~\cite{bentley2020complexity} recently gave a linear-time algorithm to find the ordering of the copies of {\tt \$} that minimizes the number of runs, but it has not been implemented and it is unclear whether it is practical for large readsets.

Another way to potentially reduce the number of runs is to remove the copies of {\tt \$} entirely, and store an auxiliary ternary vector marking which characters in the EBWT are the first and last characters in the strings.  If there are $t$ strings in the collection with total length $n$, then storing this vector takes $O (t \log (n / t) + t)$ bits (even if some of the strings are empty or consist of only one character).  As shown in Figure~\ref{fig:dollarless}, the EBWT becomes {\tt TTTTTTGTCGGGAACAAAAAATTAAAA}, with only 10 runs. The idea of replacing {\tt \$}'s with an auxiliary vector is relatively new since it originates from seeing the EBWT as a special case of Wheeler graphs~\cite{gagie2017wheeler} which are described in the next section. 

\begin{figure}[t]
\begin{center}
\begin{tabular}{rcl@{\hspace{6ex}}rcl}
&& $F$ \hspace{2.7ex} $L$ &&& $F$ \hspace{2.7ex} $L$\\
\hline
 0 & 0 & \tt ACATAT  & 14 & + & \tt GATA\ C\\
 1 & 0 & \tt ACGA\ T & 15 & 0 & \tt GATATA\\
 2 & 0 & \tt AGATAT  & 16 & 0 & \tt GATT\ A\\
 3 & $-$& \tt AGAT\ T &  17 & + & \tt GATT\ A\\
 4 & 0 & \tt AGAT\ T & 18 & 0 & \tt TACATA\\
 5 & + & \tt ATACAT  & 19 & 0 & \tt TACG\ A\\
 6 & 0 & \tt ATAC\ G & 20 & + & \tt TAGATA\\
 7 &$-$& \tt ATAGAT  & 21 & 0 & \tt TAGA\ T\\
 8 & 0 & \tt ATATAC  & 22 & 0 & \tt TAGA\ T\\
 9 & 0 & \tt ATATAG  & 23 &$-$& \tt TATACA\\
10 & 0 & \tt ATTA\ G & 24 & 0 & \tt TATAGA\\
11 &$-$& \tt ATTA\ G & 25 & 0 & \tt TTAG\ A\\
12 & 0 & \tt CATATA  & 26 & + & \tt TTAG\ A\\
13 &$-$& \tt CGAT\ A &    &\\
\end{tabular}
\caption{The matrix whose rows are the lexicographically sorted rotations of {\tt GATTA}, {\tt TTAGA}, {\tt TAGATA}, {\tt GATAC} and {\tt ATACAT}.  The EBWT is {\tt TTTTTTGTCGGGAACAAAAAATTAAAA} with 10 runs.}
\label{fig:dollarless}
\end{center}
\end{figure}

\subsection{Wheeler Graphs and XBWT}
\label{subsec:WG}

\begin{figure}[t]
\begin{center}
\includegraphics[width=.6\textwidth]{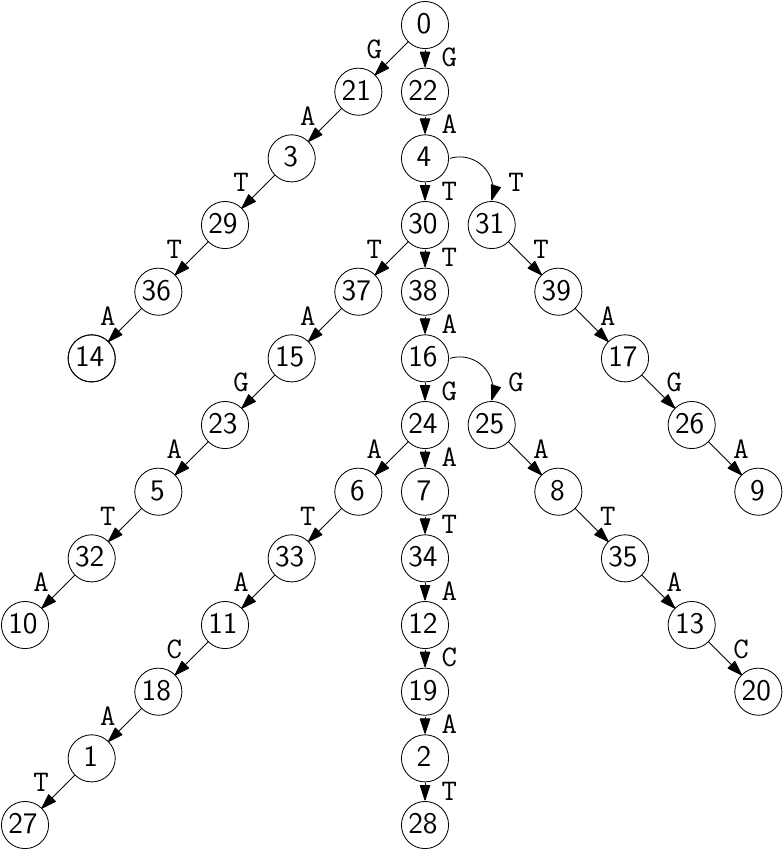}
\caption{A directed, edge-labelled tree whose vertices are labelled to show it is a Wheeler graph.  The XBWT is {\tt GGTTTTTTTTTCCCGGGGAAAAAAAAATTTTAAAAAAAA} with 7 runs.}
\label{fig:XBWT}
\end{center}
\end{figure}

Wheeler graphs were introduced by Gagie, Manzini and Sir\'en~\cite{gagie2017wheeler} as a unifying framework for several extensions of the BWT, including the EBWT, Ferragina et al.'s~\cite{ferragina2009compressing} eXtended BWT (XBWT) for labelled trees, Bowe, et al's.~\cite{bowe2012succinct} index (BOSS) for de Bruijn graphs, and Sir\'en et al.'s~\cite{siren2014indexing} Generalized Compressed Suffix Array (GCSA) for variation graphs.  A directed edge-labelled graph is a Wheeler graph if there exists a total order on the vertices such that
\begin{itemize}
\item vertices with in-degree 0 are earliest in the order;
\item if $(u, v)$ is labelled $a$ and $(u', v')$ is labelled $b$ with $a \prec b$, then $v < v'$;
\item if $(u, v)$ and $(u', v')$ are both labelled $a$ and $u < u'$ then $v \leq v'$.
\end{itemize}
Figure~\ref{fig:XBWT} shows an example of a Wheeler graph with a valid order on the vertices.  {The ordering is obtained by lexicographically sorting the strings spelling the labels in the upward path from each vertex to the root where the ties are broken deterministically (following an arbitrary order on the branches). For example, vertex 0 has upward path $\varepsilon$, vertex 3 has upward path {\tt AG}, vertex 30 has upward path {\tt TAG} and so on.} Notice that for directed acyclic graphs such as trees, such order on the vertices can be computed quickly with an adaptation of the doubling algorithm~\cite{doubling_algorithm}.

{Once we have a valid order, the standard representation of a Wheeler graph is defined considering the vertices in that order and listing the labels on the outgoing edges of each vertex. In addition, for each vertex we represent its out-degree and in-degree in unary thus obtaining two additional binary arrays. For example, for the graph in Figure~\ref{fig:XBWT} the first five vertices have outgoing edges labelled \texttt{GG T T T TT}, so the label array starts with {\tt GGTTTTT}$\cdots$ and the out-degree bit-array starts with {\tt 001010101001}$\cdots$. This simple representation, combined with {\sf rank} and {\sf select} primitives, supports efficient search and navigation operations on Wheeler graphs.} We refer the reader to Prezza's~\cite{prezza2021subpath} recent survey for a discussion of Wheeler graphs and related results.

Note that the graph in Figure~\ref{fig:XBWT} is a labelled tree: indeed its Wheeler Graph representation is equivalent to the output of the XBWT~\cite{ferragina2009compressing} applied to the same tree (details in the full paper). For clarity of presentation in the following we will still refer to the EBWT and XBWT even if they are both special cases of Wheeler graphs.

\section{Our contribution} \label{sec:main}

{Figure~\ref{fig:XBWT} can be seen as a representation of a ``genome''  {\tt GATTAGATACAT} and of five ``reads'' {\tt GATTA}, {\tt TTAGA}, {\tt TAGATA}, {\tt GATAC} and {\tt ATACAT} extracted, without errors, from it. Starting with the vertex with rank 28, corresponding to the last symbol of the ``genome'', and navigating the tree we are able to recover all the individual strings.  Notice however, that the XBWT has only 7 runs while the BWT of the ``genome'' and the EBWT of the ``reads'' in Figure~\ref{fig:dollars} have 8 and 19 runs, respectively. The EBWT without {\tt \$} of the reads alone in Figure~\ref{fig:dollarless} has 10 runs. We refer the reader to Giuliani et al.'s~\cite{10.1007/978-3-030-67731-2_18,giuliani2019dollar} recent papers for a discussion of the impact of the {\tt \$} and of the direction of the string on the number of runs in the BWT. The following theorem shows that the example in  Figure~\ref{fig:XBWT} is not a coincidence: if the ``reads'' have no errors and they are appended to the reference in the proper positions, then the XBWT has the same number of runs as the BWT of the {\it reverse} of the ``genome''.}

\begin{theorem}\label{thm:XBWT}
Suppose we sample substrings from a string and we form a labelled tree by grafting (appending) the substrings in the same position they were sampled so that all edge labels at the same depth are equal. Then the XBWT of the tree has the same number of runs as the BWT of the reverse of the string.
\end{theorem}

\begin{proof} 
Consider the tree shown in Figure~\ref{fig:XBWT}.  The tree satisfies the hypothesis of Theorem~\ref{thm:XBWT} since it was obtained by sampling some substrings from {\tt GATTAGATACAT} and then grafting them onto it such that all the edge labels at the same depth are equal  (so a horizontal line always hits edges with only the same label). Clearly, all the labels at the same depth not only are equal, but they have the same upward-path label, which is the prefix preceding the corresponding character in the string. {Since the XBWT is built by sorting labels according to the string spelled by their upward path, we see that each symbol of the original string will be adjacent to all reads symbols at the same horizontal level, and that all such symbols are identical. Finally, observe that also in the BWT of the reverse of the string symbols are sorted according to the prefix preceding them; hence the XBWT can be obtained by replacing each symbol in the BWT, except the {\tt \$}, by a run of the same symbol and the thesis follows.}
\end{proof}

Figure~\ref{fig:XBWT} and Theorem~\ref{thm:XBWT} suggest a new way to compress and index readsets: graft the reads onto a fully or partially assembled genome, or a reference genome if need be, and store the XBWT of the resulting tree.  We note that, although assembly-free indexing is a more general problem, indexing assembled reads is still of practical interest~\cite{dolle2017using}.  Many readsets have coverage of 30x or even 50x, which makes them extremely large but should also make run-length compression practical on the XBWTs.
If we want to index readsets from several individuals, we can simply graft the reads onto the appropriate assembled genomes and compute the XBWT of the forest, which is also a Wheeler graph.

{Theorem~\ref{thm:XBWT}, provides an extremely good estimate of the number of runs of the XBWT, but it holds under the unrealistic assumption that the reads have no errors. However, we can take advantage of the fact that sequencing by synthesis has an asymmetric error profile: errors are much more likely at the end of a read than at the beginning. The following result shows that errors at the end of the reads have a limited impact to the overall number of runs in the XBWT.}

\begin{theorem}\label{thm:XBWTerr}
{In the hypothesis of Theorem~\ref{thm:XBWT} suppose that the sampled substrings may differ from the reference string and that the average distance from {\em first} difference (insertion, deletion, or substitution) to the {\em end} of the substring is $\delta$. Then, with respect to Theorem~\ref{thm:XBWT} the XBWT of the tree will have at most $2\delta$ additional runs per substring.}
\end{theorem}

\begin{proof}
{Consider a single substring of length $\ell$ in which the distance between the first difference and the end of the substring is $d$ (we assume $d=0$ if there are no differences). Reasoning as in the proof of Theorem~\ref{thm:XBWT}, we see that the first $\ell-d$ symbols of the substring will end up in the same run as the corresponding symbol of the reference string (the one at the same depth in the tree). Each of the other $d$ symbols will, in the worst case, end in the middle of a run of a different symbol thus creating two additional runs. Summing this additional runs over all substrings we get a total number of additional runs upper bounded by $2\delta$ runs per substring.}
\end{proof}

To guarantee that most of the errors are at the end of the reads, we propose to build two trees: one for the assembled genome and one for its reverse complement. Having two trees means we do not have to reverse and complement half the reads before grafting them onto a single tree: the reversal of the string would be problematic in view of Theorem~\ref{thm:XBWTerr} since it would move an error from the end of the read to its front. We can build two trees with a small additional cost since the alignment algorithm will tell us whether each read aligns to the reference or to its reverse complement.

Assuming our scheme guarantees an improvement in compression we want to be sure the resulting index is also efficient. Prezza~\cite{prezza2021locating} recently showed how to generalize Gagie, Navarro and Prezza's~\cite{gagie2020fully} results about fast locating from run-length compressed BWTs to run-length compressed XBWTs, at the cost of storing the trees' shapes, which takes a linear number of bits.  For trees with far more internal vertices than leaves, however, it is relatively easy to support fast locating in small space, as a corollary of the following theorem.

\begin{theorem}
\label{thm:locating}
Let $G$ be a Wheeler graph and $r$ be the number of runs in a Burrows-Wheeler Transform of $G$, and suppose $G$ can be decomposed into $\upsilon$ edge-disjoint directed paths whose internal vertices each have in- and out-degree exactly 1.  We can store $G$ in $O (r + \upsilon)$ space such that later, given a pattern $P$, in $O (|P| \log \log |G|)$ time we can count the vertices of $G$ reachable by directed paths labelled $P$, and then report those vertices in $O (\log \log |G|)$ time per vertex.
\end{theorem}

\begin{corollary}
\label{cor:locating}
Let $T$ be a labelled tree on $n$ vertices obtained by grafting reads onto their assembled genome as described.  Let $r$ be the number of runs in the XBWT and let $t$ be the number of reads.  We can store $T$ in $O (r + t)$ words of space such that later, given a pattern $P$, in $O ((|P| + k) \log \log n)$ time we can report all the $k$ vertices reachable by paths labelled~$P$.
\end{corollary}

We sketch a proof of Theorem~\ref{thm:locating} in~\ref{app:locating}, although we omit the details because, at least when dealing with short reads, it may be more practical just to descend until we reach a branching node (in which case the pattern is in the assembled genome, not in a read) or a leaf. We have not yet considered carefully whether Nishimoto and Tabei's~\cite{nishimoto2020faster} faster locating can be applied to improve Theorem~\ref{thm:locating} or Corollary~\ref{cor:locating}.

Before we concentrate on optimizations we should consider two basic questions: are our XBWTs for readsets significantly smaller than their EBWTs in practice and, if so, how can we build them efficiently? Theorem~\ref{thm:XBWTerr} offers some guarantees of compression, but to test how our idea works in practice in Section~\ref{sec:practice} we build the XBWT and EBWT for a real, high-coverage readset and see how the numbers of runs in them compare. In Section~\ref{sec:pfp} instead we face the problem of the efficient construction of XBWTs for large datasets.

\section{XBWT via Prefix Free Parsing}\label{sec:pfp}

The problem of building the XBWT for a set of reads as described in Section~\ref{sec:main} is non trivial because the input typically consists in tens of gigabytes of data and we cannot make use of the available algorithms~\cite{JarnoSODA20,DBLP:conf/dcc/BaierBOW20} which are designed to work in RAM. However, the fact that reads are copies (possibly with errors), of portions of a relatively small reference suggests that the overall amount of information content is relatively small. Therefore we decided to compute the XBWT using the technique of Prefix Free Parsing (PFP) that has been successfully utilized for computing the BWT for large collections 
of genomes from individuals of the same species. Our implementation was done in C++ and is available on \href{https://github.com/fnareoh/Big_XBWT}{ github.com/fnareoh/Big\_XBWT}. {Note that our algorithm does not take as input a labelled tree, but rather a reference genome and a set of reads aligned to that genome (in the format of a {\tt .bam} file); the alignment implicitly defines a labeled tree as described in Section~\ref{sec:main}.}

\ignore{Since such readsets are quite large and software for building XBWTs has not been as heavily engineered as software for building BWTs and EBWTs, we first adapt prefix-free parsing (PFP) to work on readsets.  Boucher et al.~\cite{boucher2019prefix} introduced PFP as a practical technique for building BWTs for databases of genomes from individuals of the same species.}

{In the PFP construction of the BWT the input is parsed into overlapping phrases using context-triggered piecewise hashing~\cite{boucher2019prefix}. If the input contains many repetitions, the use of context-triggered hashing ensures that the parsing will contain a relatively small number of distinct phrases. The actual construction of the BWT is done using only the dictionary of distinct phrase and the parse (which describes how the dictionary phrases can be used to reconstruct the input). For repetitive datasets the dictionary and the parse fit in RAM even when the original input does not.} Unmodified, however, PFP does not work well on readsets since the phrases generated at the beginning and end of each read will likely be unique. As a result, the dictionary will be quite large and the algorithm inefficient. To prevent this, we extend the reads forward and backward so they begin and end with complete phrases. The extension is done using the symbols in the reference immediately before and after the position where the read aligns, so that the phrases are likely to be not unique (if the read has no errors the phrases will be exactly the same generated when parsing the reference). Although this technique maintains the dictionary small, the tricky part is to exclude these extensions when computing the actual XBWT.

Summing up, our implementation is divided in three main phases. In the first phase we partition the reference and the reads into phrases; the set of distinct phrases is called the {\em dictionary} and the way phrases form the reference and the reads is called the {\em parse}. We use the extension trick mentioned before, and ,if the reference and the reads are similar, the dictionary will be relatively small.   
In the second phase we compute the XBWT of the parse. Since phrases are relatively large, the number of symbols in the parse is much smaller than in the original input, so the parse fits in RAM and the computation can be done using a doubling algorithm~\cite{doubling_algorithm}. Finally, in the third phase we recover the XBWT of the input from the XBWT of the parse.  {The details of the three phases are given below.}

\subsection{Construction of the Dictionary and the Parse}

We start by scanning the reference as in the PFP BWT construction algorithm. The algorithm takes as input parameters a window size $w$, and a modulo $m$. We slide a window of length $w$ over the text, at each step computing the Karp-Rabin fingerprint~\cite{KRfingerprint} of the window. We define a terminating windows as a window with Karp-Rabin fingerprint equal to zero modulo~$m$. Terminating windows decompose the text into overlapping phrases: each phrase is a minimal substring that begins and ends with a terminating window. Note that each terminating window is a suffix of the current phrase and the prefix of the next phrase so consecutive phrases have a size-$w$ overlap. Note that defining phrases using terminating windows ensures that no phrase is a prefix (or a suffix) of another phrase, hence the name ``prefix free parsing''.

In addition to keeping track of window fingerprints, we also maintain a different hash $h(p_i)$ of the current phrase $p_i$. For simplicity in the following we assume distinct phrases always have distinct hashes, if not we detect it and crash. At the end of this scanning phase, the reference has been parsed into the (overlapping) phrases $p_1,p_2,\dots,p_z$. We build a vector $S[1,z]$ storing for each phrase $p_i$ its starting position $s_i$ in the reference and its hash $h(p_i)$.  {We also build as we go the dictionary that associate to each hash value $h(p_i)$ the corresponding phrase $p_i$  (stored as a simple string) and $occ(p_i)$ the number of occurrences of that phrase. We will later also need the length of each phrase but we don't store it explicitly, just deduce it from the string stored in the dictionary.}

%In the dictionary, when terminating a phrase we first check if the hash is already present, if not we add it and set the frequency to $1$, else we check that the phrase to save matches the phrase stored in the dictionary, if they don't we crash the program, else we just increase the frequency.

%For reasons that will become clear later on, we add $w$ NULL to the first phrase.

%{In the parse file, we separate the parse of the reference and of each read by writing $p+1$ as a separator. $p+1$ is a clear separator because the character of this temporary parse are the Karp-Rabin fingerprints of the phrases modulo $p$, thus strictly less than $p$.}
%

After parsing the reference, we process the reads one by one. 
From the file of aligned reads, we obtain both the read $r$ as a string and the position $l$ where the read aligns to the reference. 
We binary search in $S$ for the rightmost phrase $p_s$ that starts before position $l$ and for the leftmost phrase $p_e$ that ends after position $l+|r|-1$. Let $p'_s$ (resp. $p'_e$) denote the prefix (resp. suffix) of $p_s$ (resp. $p_e$) ending (resp. starting) immediately before (resp. after) position $l$ (resp. $l+|r|-1$). We define the extended read $r_{ext} = p'_s \cdot r \cdot p'_e$ where $\cdot$ here denotes string concatenation. We slide a window onto $r_{ext}$, decomposing it into phrases, as we did for the reference. Since $r_{ext}$ starts and ends with a terminating window the phrases we add while parsing $r_{ext}$ still form a prefix-free parsing.
However, as we do not want to index the whole $r_{ext}$ in the final XBWT, for each read we keep track and store to disk the starting and ending position of $r$ in $r_{ext}$.

%We then deduce for each phrase the limits of the phrase: the starting and ending positions of the characters that belong to $r$. Because we will later need this information for every phrase, we set it to the entire phrase for the phrases of the reference. Those limits are written directly to disk.

When processing the reads we continue adding the hashes of the phrases to the end parse, using a special value as separator between reads. If we parse a new phrase, we add it to the dictionary. However, as previously pointed out, the phrases coming from the extended reads are likely to be equal to phrases in the reference so we expect the dictionary not to grow significantly (the dictionary would not grow at all if all the reads were substrings of the reference). From the starting and ending position of the original read in the extended read we deduce for each phrase what characters are part of the original read (the reads without extensions) and we store a starting and ending position for each phrase.

Once all the reads have been processed, we sort the phrases in the dictionary in reverse lexicographic order and we output a new parse where each hash of phrase is replaced by its reverse lexicographic rank, the separator symbol is replaced by the number of phrases plus one. To summarize, at the end of this phase we have produced the following output files:
\begin{enumerate}
    \item \texttt{file.dict}: the dictionary in co-lexicographic order;
    \item \texttt{file.occ}: the frequency of each phrases;
    \item \texttt{file.parse}: the parse with each phrase represented by its co-lexicographic rank;
    \item \texttt{file.limits}: the starting and ending position of the original input (reads without extension) in each phrase.
\end{enumerate}

% reverse the strings in the dictionary and sort them lexicographically, thus the strings are in sorted in co-lexicographic order. Their new id is their order in the sorted dictionary and we read the temporary parse with hashes from disk to map it to the new ids. The separator $p+1$ is mapped to the total number of phrases in the dictionary plus one. We then write the dictionary and the frequency to disk.

\subsection{XBWT of the Parse}

The main goal of this phase is to construct the XBWT of the parse, using the co-lexicographic rank as meta-characters. To this end we load the parse on RAM, reconstruct its tree structure, and compute the XBWT of this tree via a doubling algorithm~\cite{doubling_algorithm}. Then, rather than storing the XBWT as is, we construct an inverted list as this structure will be more appropriate for the next phase. For each phrase $p_i$ we store the list of XBWT positions where $p_i$ appears. The size of the inverted list for $p_i$ is equal to its frequency; since frequencies were computed in the first phase, we can output the inverted list as a plain concatenation of positions.  

% So it is enough to write to disk 4 bytes per entry for a total of $4|P|$ bytes. 

In this phase we also permute the limits (the starting and ending position in the original input) of each phrase according to their order in the XBWT. This way, in the next phase, with the inverted list, we can easily access the limit of any given phrase in the parse. In this phase, we also compute and write to disk for every phrase, the list of phrases (with multiplicities) that immediately follow in the parse. This list will be used to index the characters that precede a full word. However because we only want to index the characters that are in the original input, we only add it after checking the limits. Finally, because we are not storing special characters to mark the end of a read or of the reference (as they would break runs), we construct a bit vector marking such positions and we permute it according to the XBWT order. To summarize, at the end of this phase we have produced the following output files:
\begin{enumerate}
    \item \texttt{file.dict}: the dictionary of the reversed phrases (from the first phase).
    \item \texttt{file.occ}: the frequency of each phrases (from the first phase).
    \item \texttt{file.ilist}: the inverted list of the parse.
    \item \texttt{file.xbwt\_limits}: the limits of the phrases in XBWT order.
    \item \texttt{file.xbwt\_end}: markers of the phrases where a read or reference ends in XBWT order.
    \item \texttt{file.full\_children}: for every word, the list of words that follows it.
\end{enumerate}

\subsection{Building the final XBWT}

This is the final phase where we compute the XBWT of the reference and of the readset. We start by sorting lexicographically the suffixes of the strings in the dictionary $D$. At this stage the dictionary $D$ contains the phrases reversed, so this is equivalent to sort in reverse lexicographic order the prefixes of all phrases. We ignore the suffixes of length $\leq w$ as they correspond to the terminating window which also belongs to the previous phrase.  The sorting is done by the gSACAK algorithm~\cite{LOUZA201722} which computes the SA and LCP array for the set of dictionary phrases. We scan the sorted elements of $D$, for $s$ a proper suffix, there are two cases, all the elements in $D$ which have $s$ as a proper suffix have the same preceding character, in this case we add it the correct number of times using the frequency of each phrase. In the other case, we use a heap to merge the inverted list writing the appropriate characters accordingly. Here when writing a character we first check that the suffix length is between the limits and only write it to file if it does. We also check if the character to be added is the last of its sequence (read or reference), if so output a 1 to signal the end of a sequence, else 0. When finding a suffix $s'$ that corresponds to an entire phrase, we use the children file to output the character at the start of the following phrase. At the end of this phase we have written to disk a file with the XBWT of the reference and readset as well as a bit vector marking which positions are the last character of a read or a genome. To summarize, in this phase we use \texttt{file.dict}, \texttt{file.occ}, \texttt{file.ilist}, \texttt{file.full\_children} and \texttt{file.xbwt\_limits}; all other files can be discarded. We output the XBWT in plain text as \texttt{file.bwt} and \texttt{file.is\_end} is the compressed bit vector marking the end of reads.

% Considering that we added $w$ NULL characters at the beginning of the reference, we do not forget any characters.

\section{Experiments} \label{sec:practice}

\ignore{
In this section we compare experimentally the strategies described above for reducing the number of runs when indexing a set of reads. 

As a first step we have implemented the construction of the XBWT for a set of reads together with a reference to which these reads have been aligned. In the following we call ``reference'' the sequence to which the reads are aligned: it is usually a genome assembled from those reads or a reference genome.  The more similar is the reference to the reads, the greater will be the compression, but our algorithm does not make any assumption on its content. }

In this section we present a first experimental evaluation of our XBWT-based approach for compressing a set of aligned reads and we compare it with the known methods based on the EBWT. We compare ourselves to the EBWT and not other compression tools for aligned readset as our long-term goal is to create an index and not just compression. Recall that our implementation and experimental pipeline is available on \href{https://github.com/fnareoh/Big_XBWT}{ github.com/fnareoh/Big\_XBWT}. For simplicity we compare the numbers of runs produced by the different algorithms. The actual compression depends on the algorithm used for encoding the run lengths: preliminary experiments with the $\gamma$ encoder show that the number of runs is a good proxy for measuring the actual compression. An accurate comparison of the time efficiency is left as a future work: we only compared the number of runs produced by our XBWT with the number of runs produced by the EBWT and some of its variants. Note that our implementation computes the XBWT of the reference genome and the readset (as described in the previous section), while the EBWT and its variants were applied only to the readset. We computed all EBWT variants using \href{https://github.com/lh3/ropebwt2}{{ropeBWT2}}~\cite{ropebwt2}; in addition to plain EBWT we also tested 2 heuristics that reorder the reads to reduce the number of runs in the EBWT: \href{https://github.com/shubhamchandak94/Spring/tree/reorder-only}{{Spring}}~\cite{spring} and reverse lexicographic order (RLO)~\cite{RLO}, the latter obtained using the option \texttt{-s} in ropeBWT2. Since our XBWT implementation does not use the \texttt{\$} symbol, for a fair comparison we measured the number of runs with and without the \texttt{\$} for EBWT, Spring+EBWT and RLO+EBWT (therefore ignoring for all algorithms the extra cost of implicitly encoding the ending position of each string). 
In our tests, we used the following readsets:
\begin{itemize}
    \item \textbf{E.coli} and \textbf{S.aureus} from the \href{http://bix.ucsd.edu/projects/singlecell/nbt_data.html}{{single-cell dataset}}~\cite{single-cell}, the references used are those linked on the single-cell website\footnote{\url{https://www.ncbi.nlm.nih.gov/nuccore/NC_000913}},\footnote{\url{https://www.ncbi.nlm.nih.gov/nuccore/87125858}}.
    \item \textbf{R.sphaeroides} We have HiSeq and MiSeq sequencing, raw and trimmed versions of the reads from the \href{https://ccb.jhu.edu/gage_b/}{GAGE-B dataset}~\cite{Gage-b}. The reference used is the longest contig assembled by MSRCA~v1.8.3~\cite{MSRCA} as it was the most accurate assembler according to the Gage-b companion paper~\cite{Gage-b}. We only considered the longest contig because our implementation doesn't handle forests of trees yet.
    \item \textbf{Human Chromosome 19}  We used
    as a reference Chromosome 19 from the CHM1 human assembly~\cite{chr19_chm1} and one of the HiSeq 2000 readsets\footnote{\url{https://www.ncbi.nlm.nih.gov/sra/SRX966833[accn]}} used to compute that assembly, considering only the reads that aligned with the reference.
\end{itemize}

None of those readsets are aligned, so we used \href{https://github.com/lh3/bwa}{{bwa mem}}~\cite{bwa} to align them to the chosen reference. In this preliminary experiments we discarded the reads that bwa aligned with the reverse-complement of the reference genome. As mentioned in Section~\ref{sec:main} our final prototype will build an XBWT of the tree with the reference and of the tree of the reversed-complemented reference. In Table~\ref{tab:stats_datasets}, we present statistics on the readsets we used: those statistics where computed only on the reads that aligned forward to the reference.

% \usepackage{graphicx}
\begin{table}[t]
\resizebox{\textwidth}{!}{%
\begin{tabular}{ccccLLc}
Dataset                                                            & Number of reads & Read length & Coverage & Avg. dist. from the first sequencing err. to the end & Prop. of reads without seq. error & Error rate \\ \hline
E.coli~\cite{single-cell}                                                             & 14139182        & 100         & 304$\times$   & 13             & 57.30\%               & 0.01\%     \\
S.aureus~\cite{single-cell}                                                           & 26654420        & 100         & 927$\times$   & 7             & 88.79\%              & 0.01\%     \\
Human Chr19~\cite{chr19_chm1}                                                      & 34167479        & 100         & 57$\times$    & 15             & 71.62\%              & 0.01\%     \\ \\
R.sphaeroides~\cite{Gage-b} &&&&&\\ \hline
HiSeq raw                                                          & 166820          & 101         & 46$\times$    & 27             & 31.34\%              & 0.04\%     \\
HiSeq\ trimmed                                                     & 134207          & up to 101         & 37$\times$    & 6                 & 83.26\%              & 0.01\%     \\
MiSeq raw                                                          & 23102           & 251         & 24$\times$    & 122            & 0.25\%               & 0.15\%     \\
Miseq trimmed                                                      & 20046           & up to 251         & 20$\times$    & 29                 & 63.55\%              & 0.03\%    \\ \hspace{0.5cm}
\end{tabular}%
}
\caption{Statistics on each dataset used in the experiments. Those statistics where computed only on the reads that aligned forward to the reference. We call sequencing error (or simply error) any difference between the genome and the reads. The coverage is simply defined as the total number of base-pairs in the reads compared to the number of base-pairs in the reference. The average distance between the first sequencing error and the end of the read and the end is computed considering that for error less read this distance is 0. Note that this parameter is exactly $\delta$ in Theorem~\ref{thm:XBWTerr}.}
\label{tab:stats_datasets}
\end{table}

Preliminary experiments, not reported here, show that removing the \texttt{\$} in the EBWT (all variants) reduces the number of runs between 2.7\% and 29.2\%. Consequently, we focus our analysis on the comparison of Plain (no read reordering) EBWT (without dollars), SPRING+EBWT (without dollars),  RLO+EBWT, with and without \texttt{\$} and XBWT.

%To have an experimental confirmation of Theorem~\ref{thm:XBWTerr} we first compared the different compression on Synthetic data generated by \href{https://github.com/lh3/wgsim}{{wgsim}}. This gives us a controllable amount of sequencing error and a 15\% of the polymorphisms being insertions or deletions. We generated the synthetic data from an E.coli assembly from the \href{http://bix.ucsd.edu/projects/singlecell/nbt_data.html}{{single-cell dataset}}~\cite{single-cell} and from the \href{https://hgdownload.soe.ucsc.edu/goldenPath/hg38/chromosomes/}{{Human chromosome 19}} that was sequenced by the \href{https://www.ncbi.nlm.nih.gov/grc}{{Genome Reference Consortium}}~\cite{Genome_ref_consortium}. For both of those genomes we generated 20 Millions reads of length 100 bp and different error rates. The results are presented in Figure~\ref{fig:synthetic}, RLO+EBWT without \$ gives better compression than with \$ but our approach still outperforms both of them, even for an error rate of 2\% which is much higher than the usual error rate for short reads that is considered to be around 0.1\%\cite{schirmer2016illumina}.

% removed since includes quality scores and reads which are discarded 
%The original reads, not aligned, in fastq format occupy 6.1G.

%We run experiments on real world datasets such as E.coli and S.aureus from the \href{http://bix.ucsd.edu/projects/singlecell/nbt_data.html}{{single-cell dataset}}~\cite{single-cell}, and Aeromonas Hydrophilia from the \href{https://ccb.jhu.edu/gage_b/}{{Gage-B dataset}}~\cite{Gage-b}.

\begin{figure}
    \captionsetup[subfigure]{justification=centering}
    \centering
    \begin{subfigure}[b]{0.49\textwidth}
        \centering
        \includegraphics[width=\textwidth]{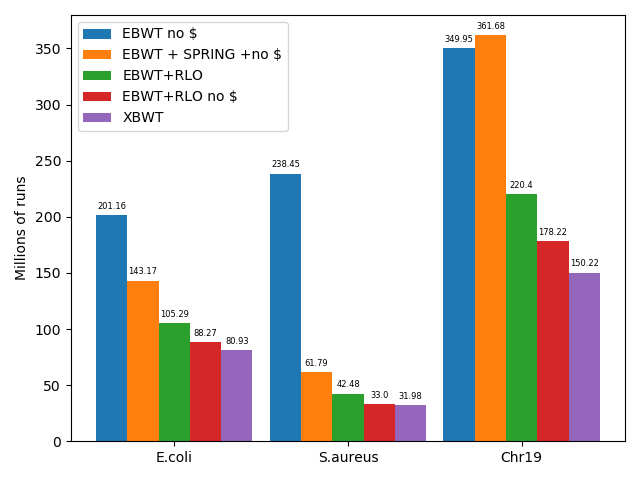}
        \caption{E.coli, S.aureus, Human Chr19}
        \label{fig:realworld}
    \end{subfigure}
    \begin{subfigure}[b]{0.49\textwidth}
        \centering
        \includegraphics[width=\textwidth]{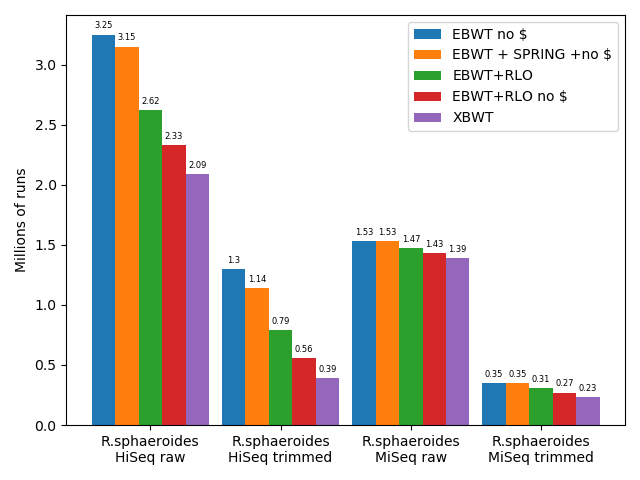}
        \caption{R.sphaeroides}
        \label{fig:sphaeroides}
    \end{subfigure}
    \caption{Comparison of run-lengths compression by RLO+EBWT with and without \$ and XBWT on various species~(\ref{fig:realworld}) and on two sequencing of R.sphaeroides (HiSeq and MiSeq) and for reads both raw and trimmed~(\ref{fig:sphaeroides}).}
    \label{fig:experiments}
\end{figure}

The results of this comparison are reported in Figures~\ref{fig:realworld} and~\ref{fig:sphaeroides}. They show that in general the plain EBWT performs worse followed by the SPRING reordering, RLO ordering with dollars then RLO ordering without dollars and finally XBWT performs best.  XBWT yields a smaller number of runs than RLO+EBWT (with or without \texttt{\$}) on all datasets, although the number is comparable on some datasets this is still a significant improvement considering that RLO+EBWT already has far less run than the EBWT baseline. On the Chr19 dataset, using RLO+EBWT-no-\texttt{\$} over plain BWT-no-\texttt{\$} (not reported in Figure~\ref{fig:realworld}) reduced the number of runs by 49\%; using the XBWT reduced the number of runs by an additional 16\%. On S.aureus and E.coli the reduction between RLO+EBWT-no-\texttt{\$} and XBWT is of only 3\% and 8\% respectively. 

The R.sphaeroides datasets are especially interesting as they involve two NGS technologies that generate reads of different lengths, different coverages, and with different error profiles. We can first notice that our method brings greater benefits on the HiSeq sequencing which has smaller reads with less errors that are located towards the end of the string. This is an experimental validation of the statement of Theorem~\ref{thm:XBWTerr}. We can also observe the effect of trimming the reads on the number of runs. On the HiSeq sequencing, trimming reduces the coverage only from 46x to 37x but yields a reduction in the number of XBWT runs by 86\%. Note that, as a result, on HiSeq trimmed, the number of XBWT runs is less than half the number of runs in plain RLO+EBWT.

%Finally, we report that the encoding of the sparse bit vector XBWT uses to mark the ending positions of substrings is relatively small as expected. For the E.coli dataset it takes 16MB while the  gamma encoding~\cite{Elias1975} of the run-lengths takes 62MB.

%However, on Hydrophylia where the reads are assembled to the reference genome we perform similarly but slightly worse, and on Hydrophilia aligned to the genome assembled by \href{https://www.bcgsc.ca/resources/software/abyss}{ABySS}~\cite{Simpson2009}, we perform significantly worse. This result is really surprising and we do not have an explanation for it yet. 

%In the future, we plan to extend this comparison to more datasets and to study the evolution of the number of runs in each method depending on the error rate and the average position of the first error.

%that does not come from a bacterial organism as genome diversity is much bigger in microbial organism~\cite{kuhnle2020efficient}.

% Synthetic reads

%All reads ?

% No reverse 

\section{Application to the JST}
\label{sec:JST}

\begin{figure}[t]
\begin{center}
\includegraphics[width=\textwidth]{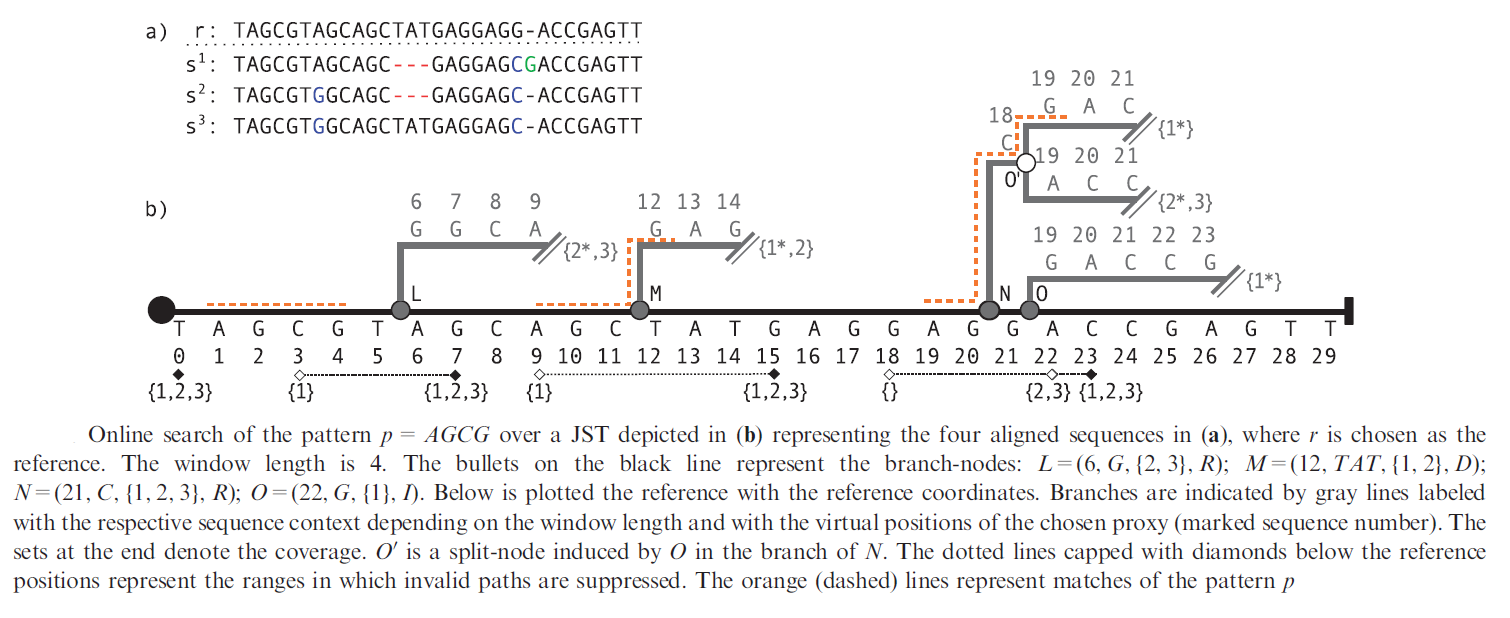}
\caption{An illustration of a JST~\cite{DBLP:journals/bioinformatics/RahnWR14}}
\label{fig:JST}
\end{center}
\end{figure}

From a certain angle, Figure~\ref{fig:XBWT} is reminiscent of Figure~\ref{fig:JST}, from Rahn, Weese and Reinert's~\cite{DBLP:journals/bioinformatics/RahnWR14} paper on their Journaled String Tree (JST).  This raises the question of whether the XBWT and JST can be used to improve the space usage of the hybrid index~\cite{ferrada2014hybrid,gagie2015searching,ferrada2018hybrid} and eventually the PanVC~\cite{valenzuela2018towards} pan-genomic read aligner, which is based on the hybrid index.

Figure~\ref{fig:JST} shows a JST supporting search for patterns of length up to 4 in four aligned sequences: the reference
\[r = \mathtt{TAGCGTAGCAGCTATGAGGAGGACCGAGTT}\]
and three others,
\begin{eqnarray*}
s^1 & = & \mathtt{TAGCGTAGCAGCGAGGAGCGACCGAGTT}\,,\\
s^2 & = &  \mathtt{TAGCGTGGCAGCGAGGAGCACCGAGTT}\,,\\
s^3 & = & \mathtt{TAGCGTGGCAGCTATGAGGAGCACCGAGTT}\,.
\end{eqnarray*}
The straight branch of the tree running along the bottom of the figure is labelled with $r$, and the other branches indicate places where the other sequences differ from $r$.  The other branches end just before a window of size 4 sliding over their sequences matches an aligned window of size 4 sliding over $r$.  For example, the first branch ends at position 9 because a sliding window of length 4 over positions 7 to 10 of sequences $s^2$ and $s^3$ (that is, containing the characters in columns 7 to 10 and the rows for $s^2$ and $s^3$ in the alignment shown at the top right in the figure), matches a sliding window of length 4 over positions 7 to 10 in $r$ (that is, containing the characters in columns 7 to 10 and the row for $r$ in the alignment).

Suppose we are looking for the pattern $p = \mathtt{AGCG}$: considering the circle at the left as the root, $p$ occurs 3 times as a substring (marked in orange) of root-to-leaf paths, and we can find those occurrences using a depth-first traversal of the tree.  Since the sequences are similar, such a traversal is faster than running a sliding window over each sequence separately.  If we find an occurrence of $p$ in the tree that ends at a node not in the branch for $r$, then we have found occurrences in each of the sequences labelling the leaves in that node's subtree.  If we find an occurrence of $p$ in the branch for $r$, then we have found occurrences in $r$ and possibly other sequences.  Unfortunately this case is not illustrated in the figure, but if we were looking for {\tt GTAG} then the occurrence at position 4 in $r$ would have a corresponding occurrence in $s^1$ but not in $s^2$ or $s^3$; this is shown by the dashed line between 3 and 7, with $\{1\}$ at the left end indicating that $s^1$ matches $r$ between 3 and 7 and $\{1, 2, 3\}$ indicating that $s^1$, $s^2$ and $s^3$ all match $r$ from 7 onward (until the next such interval starts at 9).

The hybrid index is conceptually similar to the JST, but the former is an index and the latter performs pattern matching by scanning the tree sequentially.  To build the hybrid index supporting search for patterns of length up to 4 in $r, s^1, s^2, s^3$, we first build a string kernel consisting of $r$ and substrings from $s^1, s^2, s^3$ that contain all the characters within distance 3 of variations from $r$, all separated by copies of a special symbol {\tt \$}:
\[\mathtt{TAGCGTAGCAGCTATGAGGAGGACCGAGTT\$CGTGGCA\$AGCGAG\$GAGCGACC\$GAGCACC}\,.\]
Any substring of length at most 4 of the the four sequences $r, s^1, s^2, s^3$ is a substring of the string kernel, and any substring of length at most 4 of the string kernel that does not include a copy of {\tt \$} is a substring of at least one of those sequences.  We then build an FM-index for the string kernel, with auxiliary data structure that allow us to quickly map occurrences of a pattern in the string kernel to occurrences in the sequences. 

It seems interesting that the string kernel for the four sequences in Figure~\ref{fig:JST} has more characters than the JST: on top of $r$, the string kernel has a substring {\tt \$CGTGGCA} and the JST has a branch labelled {\tt GGCA}; the string kernel has {\tt \$AGCGAG} and the JST has {\tt GAG}; the string kernel has {\tt \$GAGCGACC} and the JST has {\tt CGAC} and {\tt GACCG} (a tie in this one case); the string kernel has {\tt \$GAGCACC} and the JST has {\tt CACC} (with the first {\tt C} shared with the branch ending {\tt CGAC}).  This difference is because the string kernel stores copies of the characters both before and after variation sites, whereas the JST stores copies only of the characters after them.  If we build an index using the XBWT of the JST, therefore, it may be smaller than the hybrid index while having the same basic functionality.  We leave exploring this possiblity as future work.

%added[id=GG]{To conclude, in this paper we propose to use assembled or partially assembled genome to construct a more compressed index of its readset. For this index we give theoretical bounds on space and query time as well a theoretical compression if the reads are error-less. We then propose a way to construct the index in practice so it may scale to large datasets and start evaluating the compression on real datasets. However a lot remains to be done to develop a practical tool. First, implementing the query operation and analysing in details the impact of false positives. Second, adding support for forest of trees, to have both forward and backward reads  and to store several readset in a unique index. Third, carefully measuring the RAM consumption and scalability compared to other XBWT construction methods. Finally we would have to optimize the implementation to exploit specific sequencing information such as pair-ended reads and evaluate the lost of information we have by not indexing unaligned reads and evaluate the index compression on third generation sequencing readsets.}

\appendix

\section{Proof Sketch for Theorem~\ref{thm:locating}}
\label{app:locating}

\ignore{Gagie, Manzini and Sir\'en~\cite{gagie2017wheeler} defined a Wheeler graph to be a directed multigraph whose edges are labelled with characters from a totally-ordered alphabet and whose vertices can be totally ordered such that those with in-degree 0 precede those with positive in-degree and, for any pair of edges $e = (u, v)$ and $e' = (u', v')$ labelled $a$ and $a'$ respectively,
\begin{itemize}
\item if $a \prec a'$ then $v < v'$,
\item if $a = a'$ and $u < u'$ then $v \leq v'$.
\end{itemize}
We call such an order on the vertices a Wheeler order.}

Let $G$ be a Wheeler graph with the vertices sorted according to the permutation~$\pi$.  A Burrows-Wheeler Transform (BWT) of $G$ according to $\pi$ is a permutation of $G$'s edge labels such that, for any pair of edges $e = (u, v)$ and $e' = (u', v')$ labelled $a$ and $a'$ respectively, if $u < u'$ then $a$ precedes $a'$ in that permutation.  For convenience, we assume that the labels of each vertex's out-edges appear in the order in $\pi$ of their destinations.  Notice there may be many BWTs for $G$ because it may have many permutations $\pi$ satisfying the Wheeler graph conditions.

Let $B$ be a BWT of $G$ according to $\pi$.  By the definition of a Wheeler graph, for any pattern $P$ over the alphabet of edge labels, the vertices reachable by directed paths labelled $P$ form an interval in $\pi$.  Moreover, if we store a rank data structure for $B$ and partial sum data structures for the frequencies of the distinct edge labels and the vertices' in- and out-degrees, then given $P$ we can find its interval in $O (|P| \log \log |G|)$ time.  Let $r$ is the number of runs (i.e., maximal non-empty unary substrings) in $B$ and suppose $G$ can be decomposed into $\upsilon$ edge-disjoint directed paths whose internal vertices each have in- and out-degree exactly 1.  Then these data  structures take a total of $O (r + \upsilon)$ space, measured in words.

Let $D$ be a such decomposition of $G$ and $n$ be the number of vertices in $G$, and assume the vertices are assigned numeric identifiers from 0 to $n - 1$ such that if $(u, v)$ is an edge and neither $u$ nor $v$ is an endpoint of a path in $D$, and $u$ has identifier $i$, then $v$ has identifier $i + 1$.  Notice these identifiers are not necessarily the vertices' ranks in $\pi$.  For convenience, we assume that even though $G$ is a multigraph, the number of edges is polynomial in $n$, so $\log \log |G| = O (\log \log n)$.  We show how, still using $O (r + \upsilon)$ space, after we have found the interval for $P$ we can then report the vertices in it using $O (\log \log n)$ time for each one.

We first prove a generalization of Bannai, Gagie and I's version~\cite{bannai2020refining} of Policriti and Prezza's Toehold Lemma~\cite{policriti2018lz77}, that lets us report the last vertex in the interval for $P$.  We then define a generalization of K\"arkk\"ainen, Manzini and Puglisi's $\phi$ function~\cite{karkkainen2009permuted}, that maps each vertex's identifier to the identifier of its predecessor in $\pi$.  Finally, we give a generalization of a key lemma behind Gagie, Navarro and Prezza's $r$-index~\cite{gagie2020fully}, that lets us compute our generalized $\phi$ function with $O (r + \upsilon)$-space data structures.  Combined, these three results yield a generalized $r$-index for Wheeler graphs.

\subsection{Generalized Toehold Lemma}
\label{subsec:toehold}

For any pattern $P [0..m - 1]$, the interval for the empty suffix $P [m..m - 1]$ of $P$ is all of $\pi$, because every vertex is reachable by an empty path.  Assume we have found the interval $\pi [s_{i + 1}, e_{i + 1}]$ for $P [i + 1..m - 1]$ and now we want to find the interval $\pi [s_i, e_i]$ for $P [i..m - 1]$.  With the partial sum data structure for the vertices' out-degrees, in $O (\log \log n)$ time we can find the interval in $B$ containing the labels of the edges leaving the vertices in $\pi [s_{i + 1}, e_{i + 1}]$.

By the definition of a Wheeler graph, the edges labelled with the first and last occurrences of $P [i]$ in that interval in $B$, lead to the first and last vertices in the interval $\pi [s_i, e_i]$ for $P [i..m - 1]$.  Using the partial sum data structures for the frequencies of the distinct edge labels and the vertices in-degrees, in $O (\log \log n)$ time we can find the ranks $s_i$ and $e_i$ in $\pi$ of those first and last vertices in $\pi [s_i, e_i]$.  It follows that in $O (\log \log n)$ time we can find $\pi [s_i, e_i]$ from $\pi [s_{i + 1}, e_{i + 1}]$; therefore, by induction, we can find the interval for $P$ in $O (|P| \log \log n)$ time.  We can count the vertices in that interval in the same asymptotic time by simply returning the size of the interval.

To be able to find the identifier of the last vertex in the interval for $P$, for each edge $(u, v)$ we store $u$'s and $v$'s identifiers if any of the following conditions hold:
\begin{itemize}
\item $(u, v)$'s label $a$ is the last label in a run in $B$;
\item either $u$ or $v$ is an endpoint of a path in $D$;
\item the vertex that follows $u$ in $\pi$ has out-degree 0.
\end{itemize}
We store a select data structure for $B$, a bitvector marking the labels $a$ in $B$ for whose edges $(u, v)$ we have $u$'s and $v$'s identifiers stored, and a hash table mapping the position in $B$ of each marked label $a$ to the identifiers of its edge's endpoints.  This again takes a total of $O (r + \upsilon)$ space.

By querying the rank data structure, the select data structure, the bitvector and the hash table in that order, we can find the identifier of the vertex reached by the edge labelled by the last copy of $P [m - 1]$ in $B$.  By the definition of a Wheeler graph, this is the last vertex in the interval $\pi [s_{m - 1}, e_{m - 1}]$ for $P [m - 1]$.  Assume we have found the interval $\pi [s_{i + 1}, e_{i + 1}]$ for $P [i + 1..m - 1]$ and the identifier of the last vertex $u$ in that interval, and now we want to find the interval $\pi [s_i, e_i]$ for $P [i..m - 1]$ and the identifier of the last vertex $v$ in that interval.  We can find $\pi [s_i, e_i]$ as described above, so we need only say how to find $v$'s identifier.

With the partial sum data structure on the vertices' out-degree and the rank data structure, in $O (\log \log n)$ time we can check whether $u$ has an outgoing edge labelled $P [i]$.  If it does then, of all its out-edges labelled $P [i]$, the one whose label appears last in $B$ goes to $v$.  By our assumption of how the vertices are assigned their identifiers, if neither $u$ nor $v$ are endpoints of a path in $D$, then $v$'s identifier is $u$'s identifier plus 1.  If either $u$ or $v$ is an endpoint of a path in $D$, then we have $v$'s identifier stored and we can use the hash table to find it from the position in $B$ of the last label $P [i]$ on one of $u$'s out-edges, again in $O (\log \log n)$ time.

If $u$ does not have an outgoing edge labelled $P [i]$ then we can use the rank data structure to find the last copy of $P [i]$ in $B$ that labels an edge leaving a vertex in $\pi [s_{i + 1}, e_{i + 1}]$.  By the definition of a Wheeler graph, this edge $(u', v)$ goes to $v$.  Unlike in a BWT of a string, however, its label may not be the end of a run in $B$: $u$ could have out-degree 0, $u'$ could immediately precede $u$ in $\pi$ and the last of its outgoing edges' labels in $B$ could be a copy of $P [i]$, and the first label in $B$ of an outgoing edge of the successor of $u$ in $\pi$ could also be a copy of $P [i]$.  This is why we store $v$'s identifier if the vertex that follows $u'$ in $\pi$ has out-degree 0.  If $(u', v)$'s label is the end of a run in $B$, of course, then we also have $v$'s identifier stored.  In both cases we use $O (\log \log n)$ time, so from the interval $\pi [s_{i + 1}, e_{i + 1}]$ for $P [i + 1..m - 1]$ and the identifier of the last vertex $u$ in that interval, in $O (\log \log n)$ time we can compute the interval $\pi [s_i, e_i]$ for $P [i..m - 1]$ and the identifier of the last vertex $v$ in that interval.  Therefore, by induction, in $O (|P| \log \log n)$ time we can find the interval for $P$ and the identifier of the last vertex in that interval.

\begin{lemma}
\label{lem:toehold}
We can store $G$ in $O (r + \upsilon)$ space such that in $O (|P| \log \log n)$ time we can find the interval for $P$ and identifier of the last vertex in that interval.
\end{lemma} 

\subsection{Generalized \texorpdfstring{$\phi$}{TEXT}}
\label{subsec:phi}

For a string $S$, the function $\phi$ takes a position $i$ in $S$ and returns the starting position of the suffix of $S$ that immediately precedes $S [i..|S| - 1]$ in the lexicographic order of the suffixes.  In other words, $\phi$ takes the value in some cell of suffix array of $S$ and returns the value in the preceding cell.  Given a pattern $P$, if we can find the interval of the suffix array containing the starting positions of occurrences of $P$ in $S$, and the entry in the last cell in that interval, then by iteratively applying $\phi$ we can report the starting positions of all the occurrences of $P$.  This is the idea behind the $r$-index for strings, which uses a lemma saying it takes only space proportional to the number of runs in the BWT of $S$ to store data structures that let us evaluate $\phi$ in $O (\log \log |S|)$ time.

We generalize $\phi$ to Wheeler graphs by redefining it such that it takes the identifier of some vertex $u$ in $G$ and returns the identifier of the vertex that immediately precedes $u$ in $\pi$.  (For our purposes here, it is not important how $\phi$ behaves when given the identifier of the first vertex in $\pi$.)  Given a pattern $P$, if we can find the interval in $\pi$ containing the vertices in $G$ reachable by directed paths labelled $P$, and the identifier of the last vertex in that interval, then by iteratively applying $\phi$ we can report the identifiers of all those vertices.

Let $J$ be the set that contains $u$'s identifier if and only if any of the following conditions hold:
\begin{itemize}
\item $u$ has out-degree not exactly 1;
\item $u$ has a single outgoing edge $(u, v)$ but $v$ has in-degree not exactly 1;
\item the predecessor $u'$ of $u$ in $\pi$ has out-degree not exactly 1;
\item $u'$ has a single outgoing edge $(u', v')$ but $v'$ has in-degree not exactly 1;
\item the edges $(u, v)$ and $(u', v')$ have different labels.
\end{itemize}
We store a successor data structure for $J$ and, if $u$'s identifier is in $J$, then we store with it as satellite data the identifier of $u$'s predecessor $u'$ in $\pi$.  Notice $u$'s identifier is in $J$ only if at least one of $u$ or $u'$ or $v$ or $v'$ is the endpoint of a path in $D$, or the label of $(u', v')$ is the the last in a run in $B$ and the label of $(u, v)$ is the first in the next run.  It follows that we can use $O (r + \upsilon)$ space for the successor data structure and have it support queries in $O (\log \log n)$ time.

Suppose we know the identifier of some vertex $u$ with identifier $i$ that is immediately preceded by $u'$ in $\pi$ with identifier $i'$.  If $u \in J$ then we have $i'$ stored as satellite data with $\succ (i) = i$.  If $u \not \in J$, then $u$ has a single outgoing edge $(u, v)$ and $u'$ has a single outgoing edge $(u', v')$ with the same label, say $a$, and $v$ and $v'$ each have in-degree exactly 1.  By our assumption on how the identifiers are assigned, the identifiers of $v$ and $v'$ are $i + 1$ and $i' + 1$ and, by the definition of a Wheeler graph, $v$ is immediately preceded by $v'$ in $\pi$.  It follows that if $i + \ell$ is the successor of $i$ then it has stored with it as satellite data $i' + \ell$, and so we can compute $\ell$ and then $i'$ in $O (\log \log n)$ time.

\begin{lemma}
\label{lem:phi}
We can store $G$ in $O (r + \upsilon)$ space such that we can evaluate $\phi$ in $O (\log \log n)$ time.
\end{lemma}

\subsection{Discussion}
\label{subsec:discussion}

Combining Lemmas~\ref{lem:toehold} and~\ref{lem:phi}, we generalize, we obtain Theorem~\ref{thm:locating}.  Since $\upsilon = 1$ for a single string labelling a simple path or cycle, Theorem~\ref{thm:locating} gives the same $O (r)$ space bound and $O (|P| + k \log \log n)$ time bound we achieve with the $r$-index for strings, where $k$ is the number of occurrences.  Nishimoto and Tabei~\cite{nishimoto2020faster} recently improved the query time of the $r$-index for strings to $O (P + k \log \log n)$ --- or optimal $O (P + k)$ for polylogarithmic alphabets --- without changing the space bound, and we conjecture this is achievable also for $r$-indexes for Wheeler graphs.

\end{document}